\documentclass[dvips]{acta}
\usepackage{supertabular,lscape,epsfig}
\usepackage{amssymb}
\usepackage{amsmath}
\DeclareSymbolFont{ppa}{OT1}{ppl}{m}{it}
\DeclareMathSymbol{\vv}{\mathalpha}{ppa}{'166}

\newfont{\hb}{rphvb at 10pt}
\newfont{\hbo}{rphvbo at 10pt}
\newfont{\bitt}{rptmbi at 12pt}
\newfont{\bits}{rptmbi at 11pt}

\SetPages{1}{1}

\SetVol{65}{2015}

\begin{document}

\newcommand{\TabCapp}[2]{\begin{center}\parbox[t]{#1}{\centerline{
  \small {\spaceskip 2pt plus 1pt minus 1pt T a b l e}
  \refstepcounter{table}\thetable}
  \vskip2mm
  \centerline{\footnotesize #2}}
  \vskip3mm
\end{center}}

\newcommand{\TTabCap}[3]{\begin{center}\parbox[t]{#1}{\centerline{
  \small {\spaceskip 2pt plus 1pt minus 1pt T a b l e}
  \refstepcounter{table}\thetable}
  \vskip2mm
  \centerline{\footnotesize #2}
  \centerline{\footnotesize #3}}
  \vskip1mm
\end{center}}

\newcommand{\MakeTableSepp}[4]{\begin{table}[p]\TabCapp{#2}{#3}
  \begin{center} \TableFont \begin{tabular}{#1} #4
  \end{tabular}\end{center}\end{table}}

\newcommand{\MakeTableee}[4]{\begin{table}[htb]\TabCapp{#2}{#3}
  \begin{center} \TableFont \begin{tabular}{#1} #4
  \end{tabular}\end{center}\end{table}}

\newcommand{\MakeTablee}[5]{\begin{table}[htb]\TTabCap{#2}{#3}{#4}
  \begin{center} \TableFont \begin{tabular}{#1} #5
  \end{tabular}\end{center}\end{table}}

\newfont{\bb}{ptmbi8t at 12pt}
\newfont{\bbb}{cmbxti10}
\newfont{\bbbb}{cmbxti10 at 9pt}
\newcommand{\uprule}{\rule{0pt}{2.5ex}}
\newcommand{\douprule}{\rule[-2ex]{0pt}{4.5ex}}
\newcommand{\dorule}{\rule[-2ex]{0pt}{2ex}}
\def\thefootnote{\fnsymbol{footnote}}
\begin{Titlepage}

\Title{Multi-Mode and Non-Standard Classical Cepheids\\
in the Magellanic System\footnote{Based on observations obtained with the
1.3-m Warsaw telescope at the Las Campanas Observatory of the Carnegie
Institution for Science.}}
\Author{I.~~S~o~s~z~y~ñ~s~k~i$^1$,~~
A.~~U~d~a~l~s~k~i$^1$,~~
M.\,K.~~S~z~y~m~a~ñ~s~k~i$^1$,~~
R.~~P~o~l~e~s~k~i$^{1,2}$,\\
P.~~P~i~e~t~r~u~k~o~w~i~c~z$^1$,~~
S.~~K~o~z~³~o~w~s~k~i$^1$,~~
P.~~M~r~ó~z$^1$,~~
\L.~~W~y~r~z~y~k~o~w~s~k~i$^1$,\\
D.~~S~k~o~w~r~o~n$^1$,~~
J.~~S~k~o~w~r~o~n$^1$,~~
G.~~P~i~e~t~r~z~y~ñ~s~k~i$^1$,\\
K.~~U~l~a~c~z~y~k$^3$~~
and~~M.~~P~a~w~l~a~k$^1$}
{$^1$Warsaw University Observatory, Al.~Ujazdowskie~4, 00-478~Warszawa, Poland\\
e-mail: (soszynsk,udalski)@astrouw.edu.pl\\
$^2$Department of Astronomy, Ohio State University, 140 W. 18th Ave., Columbus, OH~43210, USA\\
$^3$Department of Physics, University of Warwick, Gibbet Hill Road,\\ Coventry, CV4 7AL, UK}
\Received{~}
\end{Titlepage}

\Abstract{We present a sample of the most interesting classical Cepheids
selected from the OGLE collection of classical Cepheids in the Magellanic
System. The main selection criterion for this sample was the presence of
non-standard, unique pulsational properties.

The sample contains the first known double-mode Cepheid pulsating in the
second- and third-overtone modes and a large number of objects with
non-radial modes excited. We also found Cepheids revealing Blazhko-like
light curve modulation, objects ceasing pulsations or showing atypical
shapes of their light curves. Additionally, the status of several triple
mode classical Cepheids is updated based on OGLE-IV photometry extending
the time baseline to 15 years.}{Cepheids -- Stars: oscillations -- Magellanic Clouds}

\Section{Introduction}
Classical Cepheids provide a unique astrophysical laboratory for studying
stellar pulsations. The vast majority of these stars are single-periodic,
radial, funda\-men\-tal-mode (F) or first-overtone (1O) pulsators. A
relatively small fraction of classical Cepheids have two consecutive radial
modes simultaneously excited: either the fundamental and first-overtone
modes (F/1O) or the two lowest-order overtones (1O/2O). This simple picture
has been significantly modified in recent years, largely thanks to the
publication by the Optical Gravitational Lensing Experiment (OGLE)
long-term, high-quality light curves of a large number of Cepheids.

The zoo of observed pulsation modes and their combinations was extended by
pure second-overtone pulsations (2O, Udalski \etal 1999), simultaneous
oscillations in the first and third overtones (1O/3O, Soszyñski \etal
2008a), and triple-mode pulsations in the fundamental, first, and second
overtone modes (F/1O/2O) or in the first three overtones (1O/2O/3O,
Moskalik \etal 2004, Soszyñski \etal 2008a). Moreover, non-radial modes
have been firmly detected in the OGLE first-overtone Cepheids (Moskalik and
Ko³aczkowski 2008, Soszyñski \etal 2008b, 2010). These modes manifest
themselves in the form of additional, small-amplitude variability with
periods which are equal to 0.60--0.65 of the dominant (first-overtone)
pulsation periods. The latter objects provided new challenges for stellar
pulsation theory (Dziembowski 2012, Smolec and ¦niegowska, in preparation).

The OGLE long-term sky-variability survey has recently released a unique
collection of classical Cepheids in the Magellanic System (Soszyñski
\etal 2015). This is the most complete and uniform sample of these
stars, containing about 10\,000 classical Cepheids from the wide area
around the Magellanic Clouds and Magellanic Bridge. The milimagnitude
accuracy and standard {\it VI}-bands of the OGLE photometry make this
dataset very attractive in searching for and studying fine effects in
pulsating classical Cepheids. The empirical constraints resulting from the
OGLE collection data should shed new light on many aspects of stellar
pulsations.

Here we present several examples of the most interesting non-standard
classical Cepheids in the OGLE collection concentrating on those revealing
unique pulsation schemes and other phenomena closely connected with stellar
pulsations. We also update the status of several multi-mode classical
Cepheids discovered by OGLE in earlier phases of the OGLE survey
(Soszyñski \etal 2008a). Thanks to additional photometric coverage
collected during the OGLE-IV phase (2010--2015) and, thus, considerable
extension of the observing time-span, it was possible to refine the
pulsating properties of these objects.

\Section{Observations and Photometric Data}
The photometry presented in this paper was collected during the third and
fourth phases of the OGLE survey in the years 2001--2015.  Since 2010, the
1.3-m Warsaw telescope with a 32-CCD detector mosaic camera covering 1.4
square degrees with a 0.26 arcsec/pixel scale was used. The Magellanic
System Survey conducted during OGLE-IV covered $\approx650$ square degrees
around the Magellanic Clouds including the Magellanic Bridge. A typical
cadence of the main {\it I}-band variability survey was 2--4
days. Observations were supplemented with much less frequent {\it V}-band
observations carried out to secure color information. The total number of
{\it I}-band epochs varied from field to field in a range of 100 to over
750. In the {\it V}-band, from several to up to 260 epochs were
collected. More information on the technical details and observing strategy
of the OGLE-IV survey can be found in Udalski \etal (2015).

Classical Cepheids were extracted from millions of variable stars with
techniques described in Soszyñski \etal (2015). The latest OGLE collection
of classical Cepheids contains 9535 objects. Precise photometry and full
characterization of each object is available from the OGLE Internet
archive. The completeness of the sample is very high -- well over
95\%. Thus, with predicted small future updates, the OGLE collection of
classical Cepheids provides the final census of these very important
variables in the Magellanic System. For more details on the OGLE collection
the reader is referred to Soszyñski \etal (2015).

\Section{Multi-Mode Cepheids}
The Petersen diagram (a plot of the ratio between two periods against the
logarithm of the longer one) is the basic and very sensitive tool to
diagnose multi-mode pulsating stars. In Fig.~1, we present the Petersen
diagram for multi-periodic Cepheids in the Magellanic Clouds constructed
using data from the OGLE collection of classical Cepheids in the Magellanic
System (Soszyñski \etal 2015). Blue and red symbols indicate the LMC and
SMC variables, respectively. Double-mode radial pulsators are marked with
filled circles, each triple-mode Cepheid is represented by three triangle
symbols, other multi-periodic Cepheids are denoted by empty circles.
\begin{figure}[p]
\vglue-3mm
\centerline{\includegraphics[width=12.7cm]{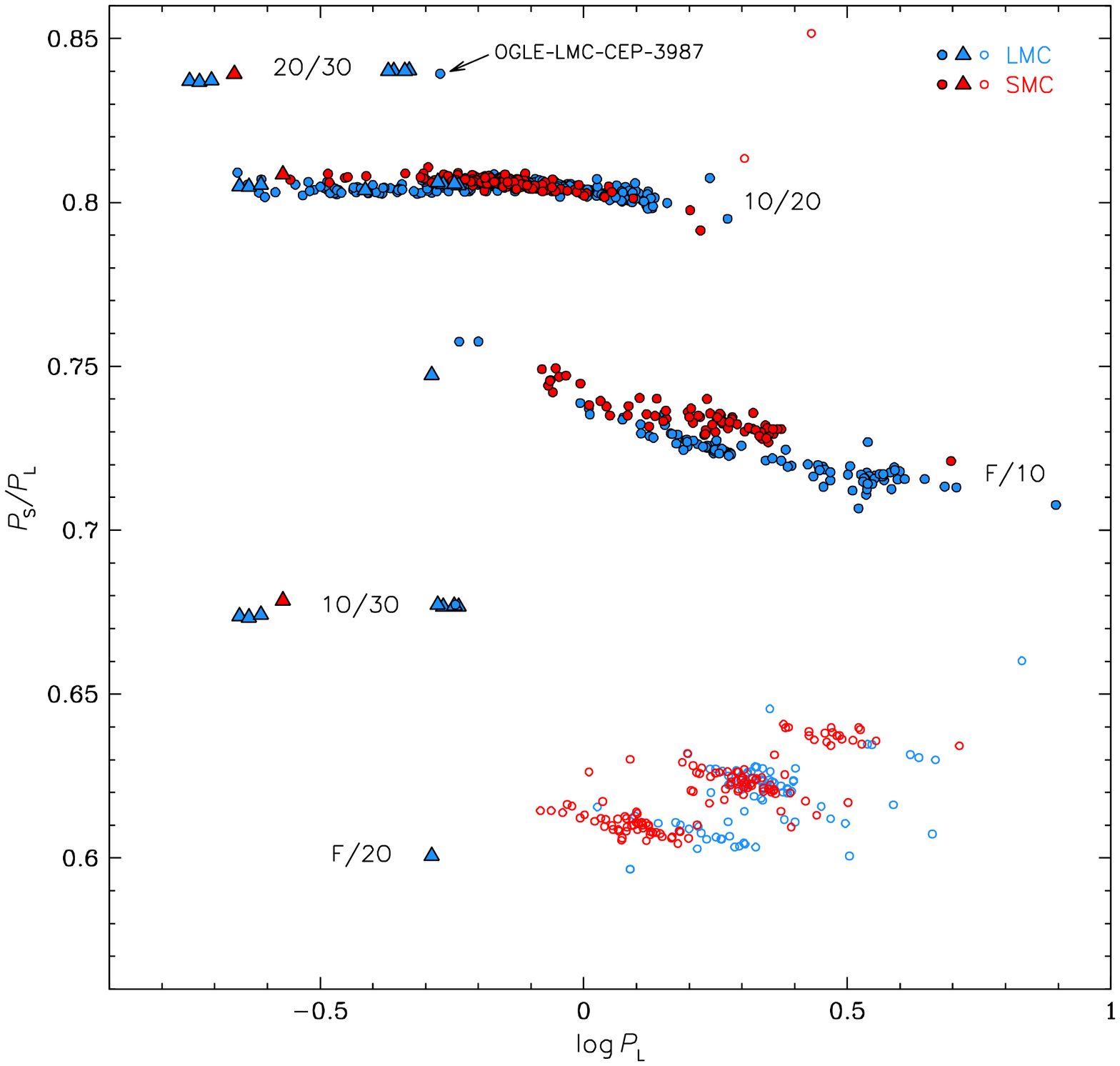}}
\vskip7pt
\FigCap{Petersen diagram for multi-mode classical Cepheids in the
Magellanic Clouds. Filled circles represent double-mode Cepheids, triangles
mark triple-mode Cepheids (three points per star) and empty circles show
other selected stars with significant secondary periods. Blue and red
symbols represent variables in the LMC and SMC, respectively.}
\vskip7mm
\centerline{\includegraphics[width=10cm]{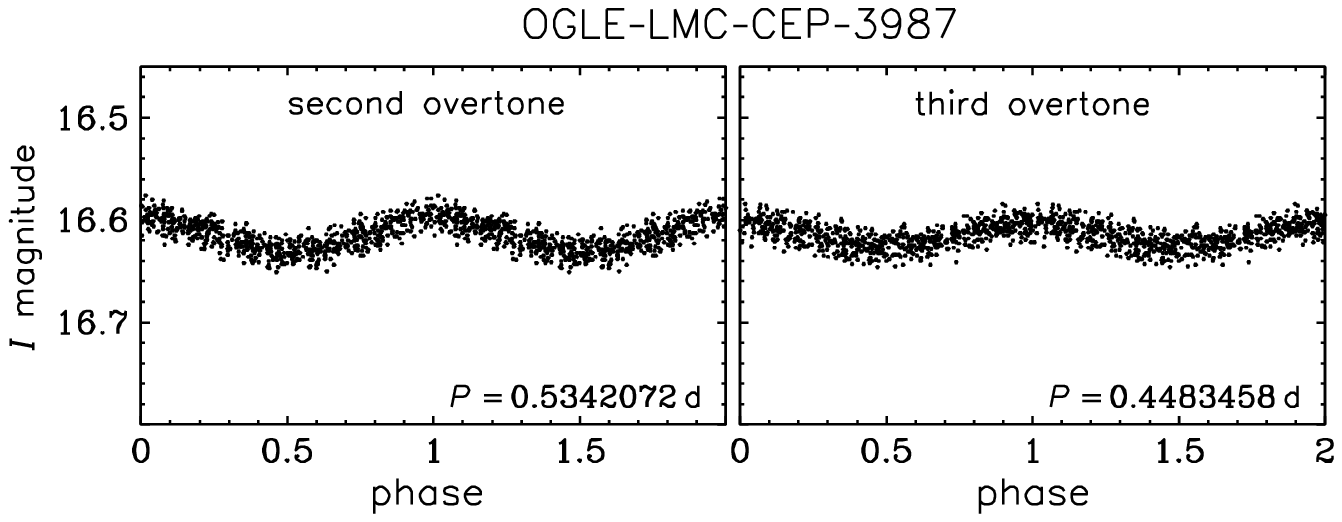}}
\vskip7pt
\FigCap{Light curve of a double-mode Cepheid OGLE-LMC-CEP-3987 pulsating in
the second and third overtones. Both modes are separated with the Fourier
technique. {\it Left} and {\it right panels} present the light curves of
the second-overtone and the third-overtone modes, respectively.}
\end{figure}

Double-mode F/1O Cepheids from both Clouds show a well-known separation in
the Petersen diagram (Beaulieu \etal 1997), while 1O/2O variables mostly
overlap. However, the largest sample of double-mode Cepheids reveals subtle
details of their distribution in the Petersen diagram, namely the shortest
period F/1O variables seem to have very similar period ratios in the LMC
and SMC, while the shortest-period 1O/2O Cepheids split up in this diagram.

\Subsection{A New Type of Double-Mode Cepheids (2O/3O)}
Among the newly identified double-mode Cepheids in the OGLE collection one
case deserves special attention. We found a unique double-mode pulsator,
OGLE-LMC-CEP-3987, which oscillates simultaneously in the second and third
overtones. To our knowledge this is the first such classical Cepheid known.

In Fig.~2, we plot the two components of its light curve, separated with
the Fourier technique. The star has a proper position in the appropriate
sequences corresponding to second- or third-overtone Cepheids in the
period--luminosity diagram plotted for the LMC Cepheids (Fig.~3). Its light
amplitudes associated with both modes are quite large and the ratio of the
periods agrees with the corresponding modes in triple-mode Cepheids
(Fig.~1). Thus, the identification of modes in OGLE-LMC-CEP-3987 seems to
be firm, despite the fact that we did not detect any combination
frequencies in the power spectrum.

\begin{figure}[t]
\vglue-5mm
\centerline{\includegraphics[width=13.9cm]{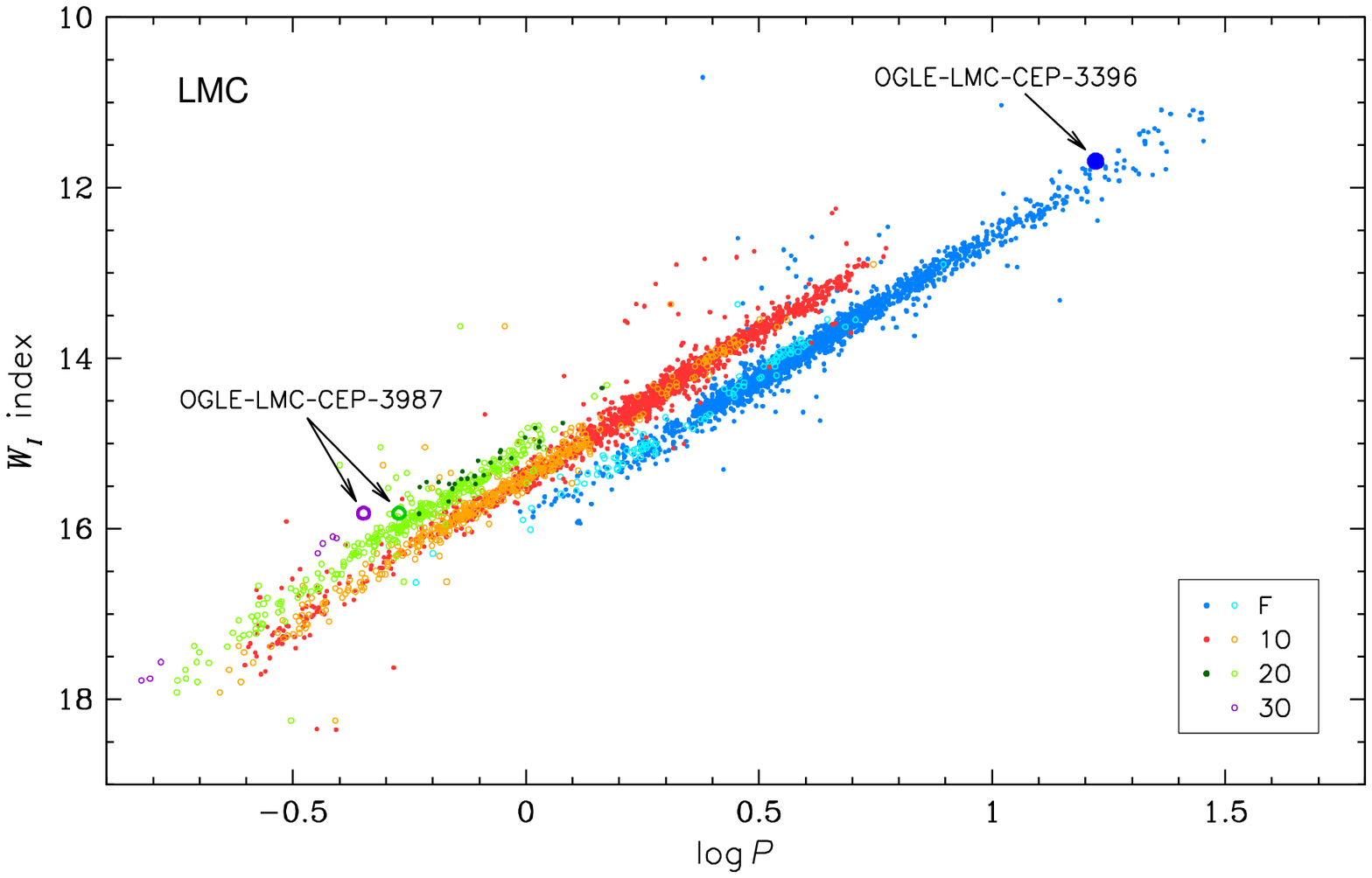}}
\vskip-9.3cm
\FigCap{Period--Wesenheit index (where $W_I=I-1.55(V-I)$) diagram for classical
Cepheids in the LMC. Blue, red, and dark-green solid circles mark F, 1O,
and 2O single-mode Cepheids, respectively. Cyan, orange, light-green and
violet empty circles represent, respectively, F, 1O, 2O, and 3O modes in
multi-mode Cepheids. Arrows show the position of the double-mode 2O/3O
Cepheid OGLE-LMC-CEP-3987 and the fundamental-mode Cepheid with a
sinusoidal light curve OGLE-LMC-CEP-3396.}
\end{figure}

\begin{figure}[t]
\centerline{\includegraphics[width=12.9cm]{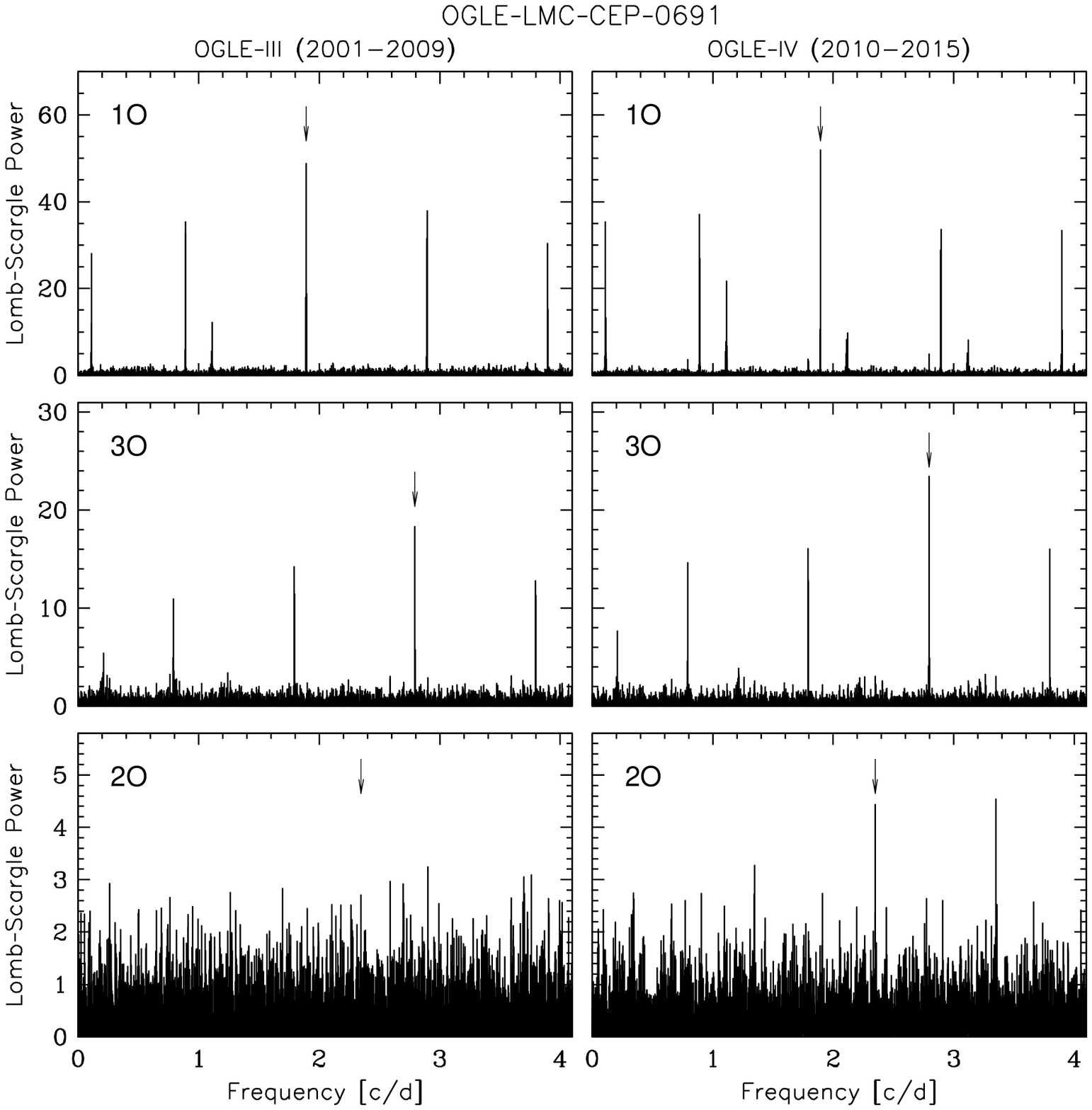}}
\vskip-4.9cm
\FigCap{Lomb-Scargle periodograms of OGLE-LMC-CEP-0691 -- a classical
Cepheid that switched from a double-mode 1O/3O to a triple-mode 1O/2O/3O
pulsator. {\it Left panels} were obtained based on the observations
collected during the OGLE-III survey (2001--2009), while the {\it right
panels} display power spectra of the same star monitored by the OGLE-IV
project (2010--2015). {\it Upper panels} show periodograms obtained for the
original light curves with the first-overtone mode indicated by the
arrows. {\it Middle panels} display power spectra after prewhitening with
the dominant (first-overtone) mode. The arrows show the frequencies of the
third overtone mode. {\it Lower panels} show the power spectra after
prewhitening with the first and the third overtone modes. Note that the
second-overtone frequency (indicated by the arrow) was hidden in noise
during the OGLE-III survey.}
\end{figure}

\Subsection{Status of Known 1O/3O Double-Mode and Triple-Mode Cepheids}
Soszyñski \etal (2008a) discovered two double-mode classical Cepheids
pulsating simultaneously in the first and third overtones (1O/3O). The
periods and period ratios of these two stars are practically the same as in
their triple-mode 1O/2O/3O counterparts. Based on that, it was suggested
that in fact double-mode 1O/3O and triple-mode 1O/2O/3O Cepheids belong to
the same class of pulsating stars. With the OGLE-IV photometry we can
confirm this suggestion, since at least one of the 1O/3O Cepheid indeed
became a triple-mode Cepheid. In OGLE-LMC-CEP-0691 we found a weak but
statistically significant signal (with full amplitude of 0.009~mag)
corresponding to the second overtone. In the OGLE-III light curve of the
same star such a signal is undetectable (Fig.~4). In the other 1O/3O
Cepheid (OGLE-LMC-CEP-1106) we still cannot detect a clear signal of the
second-overtone oscillation. We suspect it may be hidden below the OGLE
detection threshold.

On the other hand, one of the triple-mode F/1O/2O Cepheids from the
OGLE-III release (namely OGLE-LMC-CEP-0857) switched to a double-mode 1O/2O
pulsations. Very weak signal of the fundamental mode (0.006~mag) detectable
in the OGLE-III light curve became invisible in the OGLE-IV dataset. With
the OGLE-IV photometry we also do not confirm the existence of the
fundamental mode in two candidate triple-mode F/1O/2O Cepheids in the SMC
reported in the OGLE-III Cepheid sample: OGLE-SMC-CEP-1077 and
OGLE-SMC-CEP-1350. Soszyñski \etal (2010) indicated that the potential
fundamental-mode periods in both stars should be treated as uncertain
detections, since amplitudes of these variations were at the detection
limit of the OGLE-III photometry. New more precise OGLE-IV observations
prove that the fundamental-mode periods reported by Soszyñski \etal (2010)
were not real. Therefore, we presently moved both objects to the group of
double-mode 1O/2O Cepheids. Furthermore, we reclassified two double-mode
Cepheids in the LMC (OGLE-LMC-CEP-3369 and OGLE-LMC-CEP-3374) as
triple-mode pulsators and found one additional triple-mode Cepheid --
OGLE-LMC-CEP-3878. Summarizing, OGLE collection now contains nine
triple-mode Cepheids in the Magellanic Clouds: one pulsating simultaneously
in the fundamental mode, first, and second overtones (F/1O/2O) and eight
with the first three overtones (1O/2O/3O) excited. Tables~1 and~2 give
periods and amplitudes of the individual modes of triple-mode Cepheids in
the OGLE collection. It is worth mentioning that 1O/2O/3O Cepheids in the
Magellanic Clouds clearly form two groups: the shorter-period one (with
$P_{1O}<0.27$~d) and the longer-period one (with $P_{1O}>0.52$~d).

\renewcommand{\arraystretch}{1}
\MakeTableee{lcccccc}{12.5cm}{Triple-Mode F/1O/2O Cepheids in the Magellanic Clouds}
{\hline
\noalign{\vskip3pt}
\multicolumn{1}{c}{Identifier} & $P_{\rm F}$ & $A_{\rm F}$ & $P_{1O}$ & $A_{1O}$ & $P_{2O}$ & $A_{2O}$ \\
 &  [days] & [mag] & [days] & [mag] & [days] & [mag] \\
\noalign{\vskip3pt}
\hline
\noalign{\vskip3pt}
OGLE-LMC-CEP-1378 & 0.5150335 & 0.041 & 0.3849422 & 0.218 & 0.3093878 & 0.046 \\
\noalign{\vskip3pt}
\hline}
\vspace{-0.3cm}

\MakeTableee{lcccccc}{12.5cm}{Triple-Mode 1O/2O/3O Cepheids in the Magellanic Clouds}
{\hline
\noalign{\vskip3pt}
\multicolumn{1}{c}{Identifier} & $P_{1O}$ & $A_{1O}$ & $P_{2O}$ & $A_{2O}$ & $P_{3O}$ & $A_{3O}$ \\
 &  [days] & [mag] & [days] & [mag] & [days] & [mag] \\
\noalign{\vskip3pt}
\hline
\noalign{\vskip3pt}
OGLE-LMC-CEP-0691 & 0.5283428 & 0.108 & 0.4259550 & 0.009 & 0.3578421 & 0.027 \\
OGLE-LMC-CEP-1847 & 0.5795056 & 0.247 & 0.4666566 & 0.028 & 0.3921303 & 0.037 \\
OGLE-LMC-CEP-2147 & 0.5412814 & 0.265 & 0.4360338 & 0.044 & 0.3662933 & 0.070 \\
OGLE-LMC-CEP-3025 & 0.5687003 & 0.271 & 0.4582428 & 0.027 & 0.3849952 & 0.030 \\
OGLE-LMC-CEP-3369 & 0.2318730 & 0.080 & 0.1866128 & 0.070 & 0.1561395 & 0.081 \\
OGLE-LMC-CEP-3374 & 0.2222560 & 0.203 & 0.1788862 & 0.072 & 0.1497371 & 0.066 \\
OGLE-LMC-CEP-3878 & 0.2443469 & 0.238 & 0.1967768 & 0.041 & 0.1647337 & 0.080 \\
OGLE-SMC-CEP-3867 & 0.2688496 & 0.289 & 0.2173832 & 0.044 & 0.1824242 & 0.087 \\
\noalign{\vskip3pt}
\hline}

\Subsection{Non-Radial Modes in Cepheids}
Some first-overtone Cepheids show small-amplitude secondary periodicities
0.60--0.65 times shorter than the primary periods. First such stars were
detected by Moskalik and Ko³aczkowski (2008) in the OGLE
database. Soszyñski \etal (2008b, 2010) significantly increased the sample
of such stars and showed that these ``0.6 Cepheids'' form two (in the LMC)
or three (in the SMC) sequences in the Petersen diagram. The secondary
periods in these stars are likely caused by non-radial modes, but it is
still unclear which modes are responsible for this phenomenon (Dziembowski
2012).

In Fig.~1, we plot with open circles the ``0.6 Cepheids'' detected among
our overtone pulsators. In the preliminary search, we found 82 and 127 such
stars in the LMC and SMC, respectively, but more sophisticated analysis of
the OGLE light curves may reveal additional objects of that type. The vast
majority of our ``0.6 Cepheids'' pulsate in a single radial
(first-overtone) mode, but we also found additional periods in two
double-mode 1O/2O Cepheids (OGLE-LMC-CEP-4482, OGLE-SMC-CEP-2770) and in
one F/1O Cepheid (OGLE-SMC-CEP-1497).

Our investigation reveals that the ``0.6 Cepheids'' in the LMC follow three
sequences in the Petersen diagram, like their counterparts in the SMC,
although the third sequence (the highest-period-ratio one) is populated
only by a few objects. The distributions of periods of the ``0.6 Cepheids''
are clearly different in both Clouds. In particular, in the first sequence
(the lowest-period-ratio one), the LMC Cepheids have on average longer
periods than in the SMC.

\Section{Non-Standard Cepheids}
In this section we present several potentially interesting
classical Cepheids in our collection with unusual pulsating properties.

\Subsection{Blazhko-Like Effect in Classical Cepheids}
The long time span of the OGLE light curves offers the opportunity to study
non-stationary properties of pulsating stars. As an example, in Fig.~5 we
present light curves of three classical Cepheids exhibiting significant
amplitude and phase modulations, a phenomenon analogous to the Blazhko
effect in RR~Lyr type stars. Such a behavior is common in RR~Lyr stars,
while in classical Cepheids it was observed only in double-mode 1O/2O
pulsators (Moskalik and Ko³aczkowski 2009) and in a second-overtone Cepheid
V473 Lyr (Moln{\'a}r and Szabados, 2014). Our collection contains more such
cases, not only in double-mode pulsators. The upper panel of Fig.~5 shows
the light curve of OGLE-LMC-CEP-1546 -- a single-mode first-overtone
pulsator.
\begin{figure}[p]
\vglue-2mm
\centerline{\includegraphics[width=12cm]{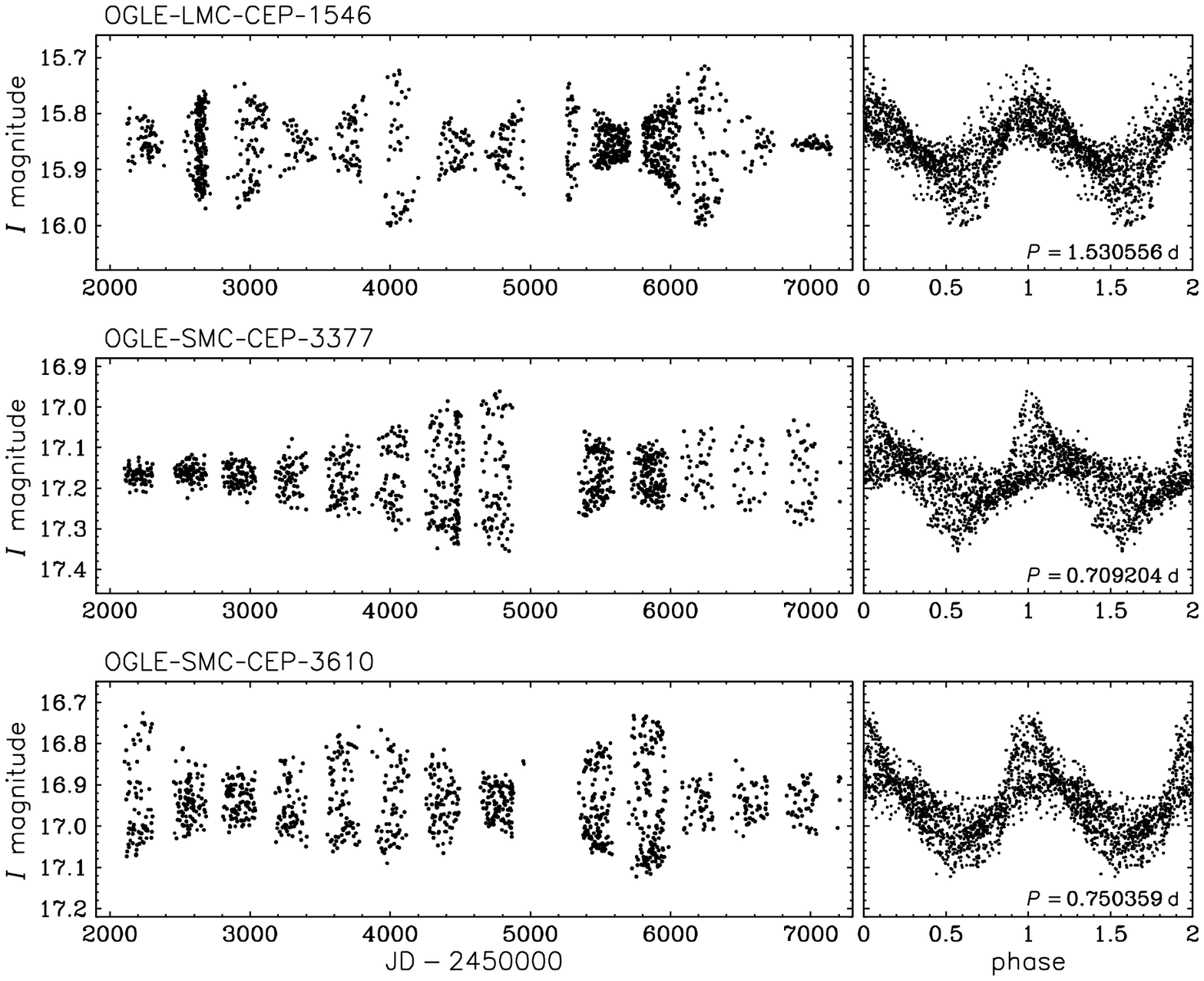}}
\vskip5pt
\FigCap{Light curves of three classical Cepheids with a Blazhko-like
effect. {\it Left panels} present unfolded light curves collected in the
years 2001--2015. {\it Right panels} show the same light curves folded with
the pulsation periods.}
\vskip4mm
\centerline{\includegraphics[width=11cm]{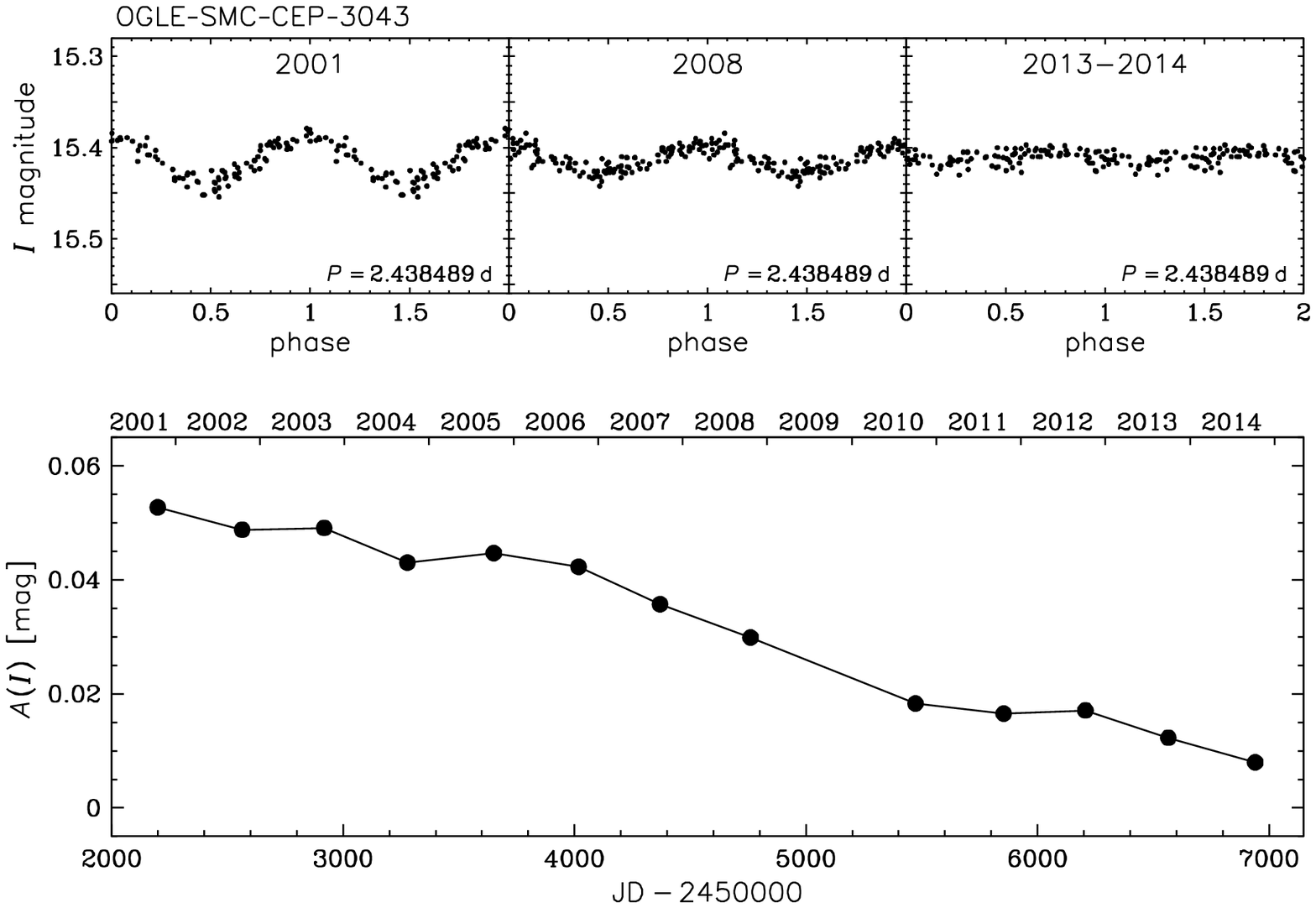}}
\vskip5pt
\FigCap{OGLE-SMC-CEP-3043 -- a first-overtone Cepheid that stopped its
pulsations. {\it Upper panels} show seasonal light curves of
OGLE-SMC-CEP-3043 in three time ranges: 2001, 2008 and 2013--2014. {\it
Lower panel} presents the changes of the pulsation amplitude through the
years.}
\end{figure}

\Subsection{Ceasing Pulsations}
OGLE-SMC-CEP-3043 is another extraordinary object found in our
collection. It is a first-overtone Cepheid observed by OGLE from the
beginning of the third phase of the project. During the 15-year-long
monitoring this star experienced a monotonic decrease of the pulsation
amplitude and currently its amplitude is below the detection limit of the
OGLE survey. In Fig.~6, we plot light curves of OGLE-SMC-CEP-3043 in three
observing seasons (in 2001, 2008 and 2013--2014). The lower panel of Fig.~6
displays changes of the amplitude through the years. Our collection
contains two more candidates for Cepheids -- OGLE-SMC-CEP-1133 and
OGLE-SMC-CEP-2081 (both of the second-overtone modes) -- that apparently
stopped their pulsations.

\Subsection{Unusual Light Curve Shape}
OGLE-LMC-CEP-3396 was discovered by Ulaczyk \etal (2013) using photometric
data collected during the OGLE-III Shallow Survey. This object reveals a
nearly sinusoidal light curve with a period of 16.68~d (Fig.~7) what is
unusual for fundamental-mode Cepheids with similar period (\cf Fig.~1 in
Soszyñski \etal 2008b). Nevertheless, despite the unusual shape of its
light curve, we confirm the classification of Ulaczyk \etal
(2013). OGLE-LMC-CEP-3396 lies in the PL relation of the fundamental-mode
classical Cepheids in the LMC (Fig.~1), the ratio of amplitudes in the {\it
V}- and {\it I}-bands ($A(V)/A(I)=1.6$) has a typical value for classical
Cepheids, and the most persuasive indication of the pulsations in this star
is a characteristic for Cepheids phase shift between the {\it V}- and {\it
I}-band light curves (Fig.~7) which excludes the possibility that this is
an ellipsoidal variable. This example illustrates potential difficulties in
the classification of variable stars in our Galaxy, when the distance and
reddening to the stars are not {\it a priori} known (see also
Pietrukowicz \etal 2015).
\begin{figure}[htb]
\centerline{\includegraphics[width=8cm]{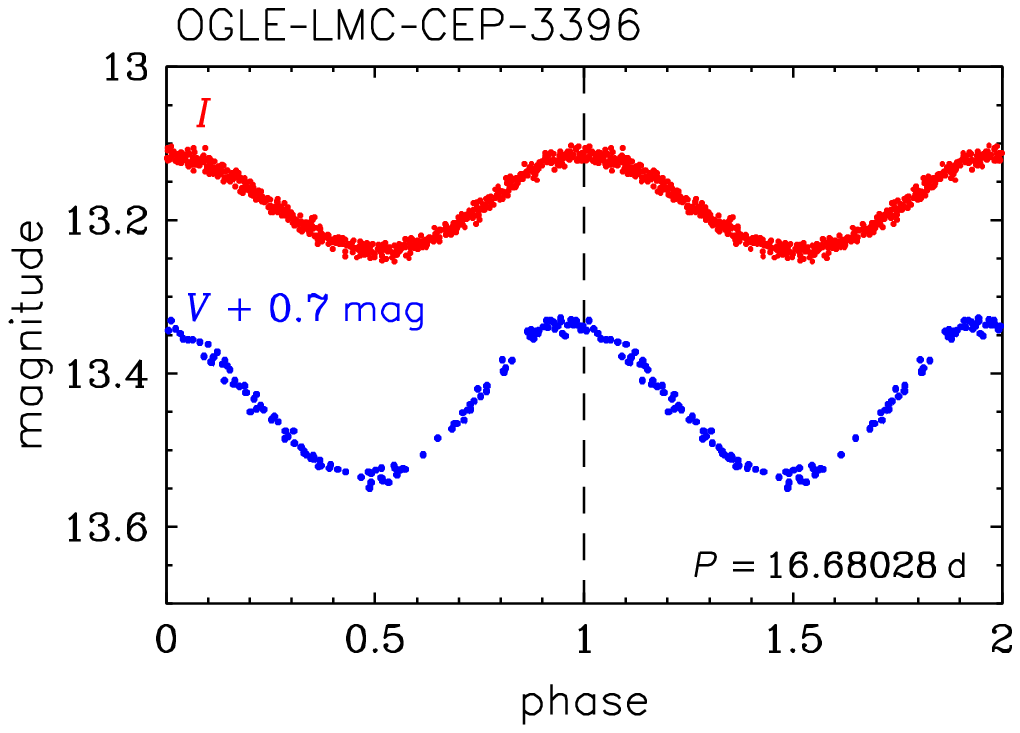}}
\vskip7pt
\FigCap{{\it I}-band (red points) and {\it V}-band (blue points) of a
fundamental-mode classical Cepheid OGLE-LMC-CEP-3396. Note the
characteristic amplitude ratio and the phase shift between the curves.}
\end{figure}

\Acknow{We would like to thank Prof.\ M.~Kubiak, OGLE project co-founder,
  for his contribution to the collection of the OGLE photometric data over
  the past years. We thank R.~Smolec for helpful comments about some multi-mode
  Cepheids. This work has been supported by the Polish Ministry of Science
  and Higher Education through the program ``Ideas Plus'' award No.
  IdP2012 000162. The OGLE project has received funding from the Polish
  National Science Centre grant MAESTRO no. 2014/14/A/ST9/00121 to AU.}

\end{document}